\documentclass[twocolumn,twocolappendix]{aastex631}

\usepackage{nccmath}

\usepackage{xfrac}

\newcommand{\nrel}{\mbox{\tiny{NR}}}

\newcommand{\jel}{\mbox{\tiny{JEL}}}
\newcommand{\lar}{\mathrm{\tiny large}}
\newcommand{\corr}{\mathrm{corr}}

\newcommand{\intern}{\mathrm{int}}

\begin{document}

\title{Finite temperature description of Fermi gases with in-medium effective mass}

\author[0000-0001-7501-0404]{Mariana Dutra}
\affiliation{Institut de Physique des 2 infinis de Lyon, CNRS/IN2P3, Universit\'e de Lyon, Universit\'e Claude Bernard Lyon 1, F-69622 Villeurbanne Cedex, France}
\affiliation{Departamento de F\'isica, Instituto Tecnol\'ogico de Aeron\'autica, DCTA, 12228-900, S\~ao Jos\'e dos Campos, SP, Brazil}

\author[0000-0002-0935-8565]{Odilon Louren\c{c}o}
\affiliation{Institut de Physique des 2 infinis de Lyon, CNRS/IN2P3, Universit\'e de Lyon, Universit\'e Claude Bernard Lyon 1, F-69622 Villeurbanne Cedex, France}
\affiliation{Departamento de F\'isica, Instituto Tecnol\'ogico de Aeron\'autica, DCTA, 12228-900, S\~ao Jos\'e dos Campos, SP, Brazil}

\author[0000-0001-8743-3092]{J\'er\^ome Margueron}
\affiliation{Institut de Physique des 2 infinis de Lyon, CNRS/IN2P3, Universit\'e de Lyon, Universit\'e Claude Bernard Lyon 1, F-69622 Villeurbanne Cedex, France}

\date{\today}

\begin{abstract}
We investigate Fermi gases at finite temperature for which the in-medium effective mass may not be constant as a function of the density, the temperature, or the chemical potential. We suggest a formalism that separates the terms for which the mass is constant from the terms which explicitly treat the correction due to the in-medium effective mass.
We employ the ensemble equivalence in infinite matter in order to treat these different terms. Our formalism is applied in nuclear matter and we show its goodness by comparing it to an exact treatment based on the numerical calculation of the Fermi integrals.
\end{abstract}

\section{Introduction}

The description of Fermi gases at finite temperature is numerically more evolved than at zero temperature, for which all expressions are analytical. Some highly accurate approximations have therefore been suggested to facilitate and accelerate the calculation of thermodynamical properties at finite temperature, see for instance Refs.~\cite{eggleton1973,Antia-1993,pols,JEL-1996-1020,Aparicio_1998,mohankumar-2005,natarajan-2001,mamedov-2012,fukushima-2014,khvorostukhin-2015,gil-2023}. Most of these approximation schemes, however, assume a constant in-medium mass for the particles, which may limit their application to describe the kinetic energy contribution of a more complex system in interaction. In other words, they surely allow us to describe a Fermi gas with a mass different from the bare mass of the associated particle, but this mass is constant and these approximations do not treat the possibility that the mass is an in-medium mass changing as a function of the thermodynamical quantities, as it can be for dense matter equation of states employed for the description of neutron star mergers or core-collapse supernovae. 

In the following, we shall distinguish the free Fermi gas (FFG), with no interaction among the particles, from the Fermi gas (FG), which incorporates an in-medium effective mass. In practice, the FG is often encountered in interacting systems and stands for the contribution of the kinetic energy term. The scope of this paper is to suggest a framework where the FFG approximations at finite temperature could still be employed to describe the properties of FGs, allowing fast calculation even at finite temperature.

We begin by defining, in Sec.~\ref{sec:exact}, the main thermodynamical quantities in terms of the Fermi-Dirac distribution, which allows exact calculations employed as a reference. In Sec.~\ref{sec:examples} we furnish some examples of functional forms for the effective mass as a function of the density, temperature, or chemical potential. An application for symmetric and asymmetric nuclear matter is given in Sec.~\ref{sec:effmass-n}, where we specifically employ the approximation suggested in Ref.~\cite{JEL-1996-1020}, hereafter called JEL, in the numerical application of our formalism. In Sec.~\ref{sec:effmass-nt} the correction induced by an effective mass depending on both, density and temperature, is derived and in Sec.~\ref{sec:effmass-cp} we consider an effective mass depending on the chemical potential. In Secs.~\ref{sec:effmass-n}-\ref{sec:effmass-cp} we employ the ensemble equivalence in the infinite matter to evaluate the different contributions to the thermodynamic quantities. Our conclusions are presented in Sec.~\ref{sec:conclusions}.

\section{Exact treatment of Fermi gas with in-medium effective mass}
\label{sec:exact}

For Fermi gas at finite temperature $T$, the thermodynamical properties such as the pressure $p$, the energy density $\epsilon$, and the entropy density $\sigma$ are expressed in terms of Fermi integral as,
\begin{align}
p &= \frac{\gamma}{6\pi^2}\int_0^{\infty}\hspace{-0.2cm}\frac{dk\,k^4}{(k^2 + 
{m^2})^{1/2}}F_D(k,\mu,T,m)\, ,
\label{eq:pressfg}
\end{align}
\begin{align}
\epsilon &= \frac{\gamma}{2\pi^2}\int_0^{\infty}\hspace{-0.2cm}dk\,k^2(k^2 + 
{m^2})^{1/2}F_D(k,\mu,T,m)\, ,
\label{eq:edenfg}
\end{align}
\begin{equation}
\sigma = -\frac{\gamma}{2\pi^2}\int_0^{\infty}dk\,k^2
[F_D\mbox{ ln} F_D + (1-F_D)\mbox{ln}(1-F_D)]\, ,
\label{eq:entdenfg}
\end{equation}
where the Fermi-Dirac distribution $F_D$ reflects the distribution of single-particle states at equilibrium, 
\begin{align}
F_D(k,\mu,T,m) = \frac{1}{e^{(\sqrt{k^2+m^2}-\mu)/T}+1},
\end{align}
with $\mu$ being the chemical potential, $m$ the mass of the particles and $\gamma$ the degeneracy factor. 
We fix $\hbar=c=k_B=1$ such that momenta, masses, and temperatures are measured in units of energy.
The density is also modified by the effect of finite temperature as,
\begin{align}
n &= \frac{\gamma}{2\pi^2}\int_0^{\infty}\hspace{-0.2cm}dk\,k^2F_D(k,\mu,T,m).
\label{eq:densityfg}
\end{align}
Note that these expressions could be generalized easily to include anti-particle contributions (with a chemical potential $\mu_{\bar{\alpha}}=-\mu_{\alpha}$).

In the case of free Fermi gas (FFG) for which there are no interactions among the particles, the mass $m$ is constant and equal to the bare mass of the particles. In the case of interacting systems however, the many-body correlations can modify the mass and generate a dependence in terms of the single-particle states, here represented by the momentum $k$, or more globally of the thermodynamical quantities such as the density, the temperature, or the chemical potential for instance. For interacting systems, the expressions obtained for the FFG are still valid to describe the kinetic energy contribution, but the in-medium mass should be properly treated. This can be done exactly by solving the integrals given in Eqs.~\eqref{eq:pressfg}-\eqref{eq:densityfg} with the proper dependence of the mass on the in-medium quantities. In the following, these calculations, using the Gauss-Legendre method with 600 Gauss points, will be used as a reference. 

The aim of the present paper is to suggest an alternative approach to the exact calculations, which could take advantage of existing fast alternatives to the numerical calculation of the Fermi integrals in \eqref{eq:pressfg}-\eqref{eq:densityfg}. In the following, we suggest corrections to the thermodynamical potentials which include the in-medium effective mass dependencies. In other words, the expressions given in our formalism separate the terms which assume a constant mass (for which analytical or semi-analytical expressions could be used) from the terms which treat explicitly the in-medium effective mass dependencies. 

\section{Examples of in-medium effective masses}
\label{sec:examples}

In this section, we give examples of in-medium effective masses which are employed in nuclear physics for illustrative purposes. Our formalism can however be applied to any kind of Fermi systems.

In nuclear systems, it is quite frequent to encounter a density-dependent effective mass $m^*$ of the following form,
\begin{equation}
\frac{m^*_q(n,\delta)}{m}=\left[1+2m\left(C_0^\tau+\tau_3C_1^\tau \delta\right)n\right]^{-1} \, ,
\label{eq:effmass:den}
\end{equation}
for $q$ indexing neutrons ($n$) and protons ($p$), and where $m$ is the nucleon bare mass (neglecting the small difference between neutrons and protons), $n=n_n+n_p$ is the total density, and $\delta=(n_n-n_p)/n$ is the isospin parameter controlling the isospin asymmetry of the medium. We remind that $\tau_3=1$ for neutrons and $-1$ for protons in our convention. The parameters $C_0^\tau$ and $C_1^\tau$ are the parameters controlling the in-medium effective mass. 

The in-medium effective mass~\eqref{eq:effmass:den} is naturally obtained from Skyrme interaction, see for instance~\cite{Tondeur-1984-297,Chabanat-1997-710,Bender-2003,Li-2018-29} and references therein, but it can also be an approximation for a more general in-medium effective mass capturing, at first order, the non-local contribution of the nuclear interaction around the Fermi level~\citep{bhf1,bhf2}.

More general dependencies of the in-medium effective mass could however be introduced along the idea of the density functional theory (DFT). An explicit temperature dependency has been introduced in Refs.~\cite{Fantina-2011,Fantina-2012} for instance, in order to describe in a phenomenological way the finite-$T$ effects originating from the dynamical mass. 

One may also imagine that the density dependence in expression~\eqref{eq:effmass:den} could be replaced by a dependence on the chemical potential. While at zero temperature the simple relation between the chemical potential and the density could be used to impose an equivalence between the two prescriptions, some differences could appear at finite-$T$. Let us suggest some forms for the in-medium effective mass with explicit dependence on the chemical potential. By employing the relation between the density and the chemical potential of a relativistic~FFG at $T=0$, $\mu_q=(k_{F_q}^2+m^2)^{1/2}$ and $\mu=(k_{F}^2+m^2)^{1/2}$, with $n_q=k_{F_q}^3/(3\pi^2)$ and $n=k_{F}^3/(3\pi^2)$, the effective mass~\eqref{eq:effmass:den} can be transformed into
\begin{align}
\frac{m}{m^*_q(\nu_n,\nu_p)} &=1+2m\Big\{\tilde{C}_0^\tau (\mu^2-m^2)^{3/2} 
\nonumber\\
&+\tau_3\tilde{C}_1^\tau \left[(\mu_n^2-m^2)^{3/2}-(\mu_p^2-m^2)^{3/2}\right] \Big\}, 
\label{eq:effmass:pot}
\end{align}
where we have $\tilde{C}_i^\tau=C_i^\tau/(3\pi^2)$ with $i=0$, $1$. 

One may also consider a simpler expression for the in-medium effective mass,
\begin{align}
\frac{m}{m^*_q(\nu_n,\nu_p)} &=1+2m\left[\bar{C}_0^\tau\mu
+\tau_3\bar{C}_1^\tau (\mu_n-\mu_p) \right], 
\label{eq:effmass:pot2}
\end{align}

and fix the constants $\bar{C}_i^\tau$ to describe the expected values of the in-medium effective mass as a function of the density and isospin asymmetry. In the future, it would be interesting to investigate further the differences induces by the functional forms~\eqref{eq:effmass:pot} and \eqref{eq:effmass:pot2}.

Other functional forms could be considered, mixing $\mu$ and $\delta$ for instance. The purpose of this section is however only illustrative since the large majority of in-medium effective masses employed in nuclear physics remains the one given by Eq.~\eqref{eq:effmass:den}. In the spirit of the DFT, one could however consider different in-medium effective masses, such as for instance the ones shown in this section. In the following, we show that our formalism could easily accommodate this variety of in-medium effective masses.

\section{Application to symmetric and asymmetric nuclear matter with density dependent in-medium effective mass}
\label{sec:effmass-n}

We first consider the symmetric nuclear matter (SM) where the isospin symmetry breaking of the bare mass is neglected. As a consequence and for spin symmetric systems, SM can be treated as a system of identical particles with degeneracy $\gamma=4$ (spin and isospin degeneracy). In asymmetric nuclear matter (AM), we still assume spin saturated matter (spin symmetry) and we break the isospin symmetry at the level of the particle composition (keeping the same bare mass). We then treat a system of neutrons and protons with degeneracy $\gamma=2$ and the SM limit could easily be recovered.

\subsection{Formalism in symmetric matter}
\label{sec:SNM}

In the case of SM for which $\delta=0$, Eq.~\eqref{eq:effmass:den} leads to $m_p^*=m_n^*=m^*$, with
\begin{equation}
\frac{m^*(n)}{m}=\left(1+2m C_0^\tau \, n \right)^{-1} \, .
\label{eq:effmass:SM}
\end{equation}
The in-medium mass depends on the density $n$, which reflects that the number of particles is controlled as in the micro-canonical ensemble (MCE) or in the canonical ensemble (CE). Since we allow thermal exchanges, a natural choice to treat density dependent in-medium effective mass is the CE, where thermodynamical quantities are functions of the density $n$ (extensive variable) and temperature $T$ (intensive variable). In the CE, the associated potential is the Helmholtz free energy per particle, namely, $f = \phi(n,T,m^*)/n$, where $\phi$ is the Helmholtz free energy density.

Since we have that, at fixed $T$,
\begin{align}
\frac{\partial X(n,m^*)}{\partial n}\Bigg|_{T} &= \frac{\partial X(n,m^*)}{\partial n}\Bigg|_{T,m^*} 
\nonumber\\
&+ \frac{\partial X(n,m^*)}{\partial m^*}\Bigg|_{T,n} \frac{\partial m^*}{\partial n}\Bigg|_{T},
\label{eq:general}
\end{align}
the FG pressure reads 
\begin{align}
p &\equiv n^2 \frac{\partial f}{\partial n}\Bigg|_{T} 
= n\frac{\partial\phi}{\partial n}\Bigg|_{T,m^*} + n \frac{\partial\phi}{\partial m^*}\Bigg|_{T,n}\frac{\partial m^*}{\partial n}\Bigg|_{T}
- \phi  \nonumber\\
&= p^* + p_{\corr},
\label{eq:pressure}
\end{align}
with 
\begin{equation}
p^* = n\frac{\partial\phi}{\partial n}\Bigg|_{T,m^*} - \phi
\label{eq:pstar}
\end{equation}
and
\begin{equation}
p_{\corr} = n \frac{\partial\phi}{\partial m^*}\Bigg|_{T,n}\frac{\partial m^*}{\partial n}\Bigg|_{T}.
\label{eq:pcorr}
\end{equation}
The term $p^*$ is the FFG pressure considering the constant mass $m^*$ instead of $m$. This notation is employed throughout the paper. In $p^*$, the density dependence of the effective mass is not considered since $m^*$ is fixed, even for the density derivative term. The last term, $p_{\corr}$, is the correction due to the density dependency of the in-medium effective mass. If the in-medium effective mass depends on the density, $p_{\corr}\ne 0$, it is important to include this correction term in the pressure, as well as in other thermodynamical quantities. The impact of $p_{\corr}$ is emphasized in the following.

Additionally, since the CE and the grand-canonical ensemble (GCE) are equivalent in infinite matter, the terms $p^*$ and $p_{\corr}$ may be computed in the most natural ensemble. The GCE is well-suited to compute $p^*$ with $\mu$ and $T$ as thermodynamical variables since the effective mass is taken as constant, and the term $p_{\corr}$ may be better calculated in the CE since it requires considering the density $n$ and the temperature $T$ with the chemical potential $\mu(n,T)$. 

Note that since the density is the thermodynamical variable of the CE, it is not impacted by the density dependent effective mass: $n_\corr=0$ and $n=n^*$.

The chemical potential can be calculated in a similar way,
\begin{equation}
\mu \equiv \frac{\partial \phi}{\partial n}\Bigg|_{T}  =  \frac{\partial\phi}{\partial n}\Bigg|_{T,m^*} + \frac{\partial\phi}{\partial m^*}\Bigg|_{T,n}\frac{\partial m^*}{\partial n}\Bigg|_{T}  
% \nonumber \\
 = \mu^* + \mu_{\corr},
 \label{eq:mu}
\end{equation}
with 
\begin{equation}
\mu_{\corr} = \frac{\partial\phi}{\partial m^*}\Bigg|_{T,n}\frac{\partial m^*}{\partial n}\Bigg|_{T} = \frac{p_{\corr}}{n}. 
\label{mucorr}
\end{equation}
Here also the term $\mu^*$ defined in Eq.~\eqref{eq:mu} is the chemical potential of the system with constant mass $m^*$. 

In the case of the entropy density, defined as
\begin{align}
\sigma = -\frac{\partial\phi}{\partial T}\Bigg|_{n},
\end{align}
there is no correction to be added if the in-medium mass is independent of $T$. We therefore obtain
\begin{align}
\sigma = \sigma^*.
 \label{eq:ssjel}
\end{align}
The energy density can be determined from the Euler relation,
\begin{align}
\epsilon &= - p + \mu \, n + T\sigma\nonumber\\
&= - (p^* + p_{\corr}) + (\mu^* + \mu_{\corr}) \, n + T\sigma^*
\nonumber\\  
&= - p^* + \mu^* \, n + T\sigma^* 
= \epsilon^*.
\end{align}
Finally, the Helmholtz free energy density is 
\begin{align}
\phi &= \epsilon - T\sigma = - p + \mu \, n
\nonumber\\
&=\epsilon^* - T\sigma^* = - p^* + \mu^* \, n 
= \phi^*,
\label{eq:freeenergy}
\end{align}
as expected, since it is the potential associated with the CE.

\subsection{Practical implementation}
\label{sec:JEL}

In this section, we show how our formalism could be employed in practice with some efficient approximations of the Fermi integrals. For our purpose, we chose the Johns-Ellis-Lattimer~(JEL) approximation~\citep{JEL-1996-1020}, which is a fast and accurate approximation for the FFG system at finite $T$. We provide a short description of this approximation in Appendix~\ref{app:jel}. 

By using the JEL approximation, it is possible to rewrite Eq.~\eqref{eq:pcorr} in a more compact form. We have
\begin{equation}
\frac{\partial\phi}{\partial m^*}\Bigg|_{n,T} = \frac{\partial\phi^*}{\partial m^*}\Bigg|_{n,T} = -\frac{\partial p^*}{\partial m^*}\Bigg|_{n,T} + n \frac{\partial \mu^*}{\partial m^*}\Bigg|_{n,T}.
\label{eq:dfdms}
\end{equation}
Using Eq.~\eqref{eq:mujel} from the JEL approximation, namely $\mu^* = \psi(n) T + m^*$, we obtain
\begin{equation}
\frac{\partial \mu^*}{\partial m^*}\Bigg|_{n,T} = 1, 
\label{dmudms}
\end{equation}
and injecting Eq.~\eqref{eq:dpjeldm} (with $m^*$ instead of $m$) into Eq.~\eqref{eq:dfdms} we find
\begin{equation}
\frac{\partial\phi^*}{\partial m^*}\Bigg|_{n,T} = \frac{1}{m^*}(\epsilon^*-3p^*),
\label{eq:dfdms2}
\end{equation}
leading to 
\begin{align}
p_{\corr} = \frac{n}{m^*}(\epsilon^*-3p^*)
\frac{\partial m^*}{\partial n}\Bigg|_{T},
\label{eq:pcorr2}
\end{align}
with $p^*$ and $\epsilon^*$ expressed from Eqs.~\eqref{eq:nodim} (with a constant mass $m^*$ replacing $m$). We emphasize here that Eq.~\eqref{eq:pcorr2} was found by explicitly using the JEL approximation. It is however interesting to note that for non-relativistic systems, since $\epsilon_{\rm int}^{\nrel*}=(3/2)p^{\nrel*}$, the expression~\eqref{eq:pcorr2} becomes 
\begin{align}
p_{\corr}^{\nrel} = -\frac{3}{2}\frac{n}{m^*}
\frac{\partial m^*}{\partial n}\Bigg|_{T}p^{\nrel*} \, .
\end{align}
This term is explicitly used in~\cite{mekjian,kuo,cp-vinas}.

\begin{figure*}[tb]
\centering
\includegraphics[scale=1.0]{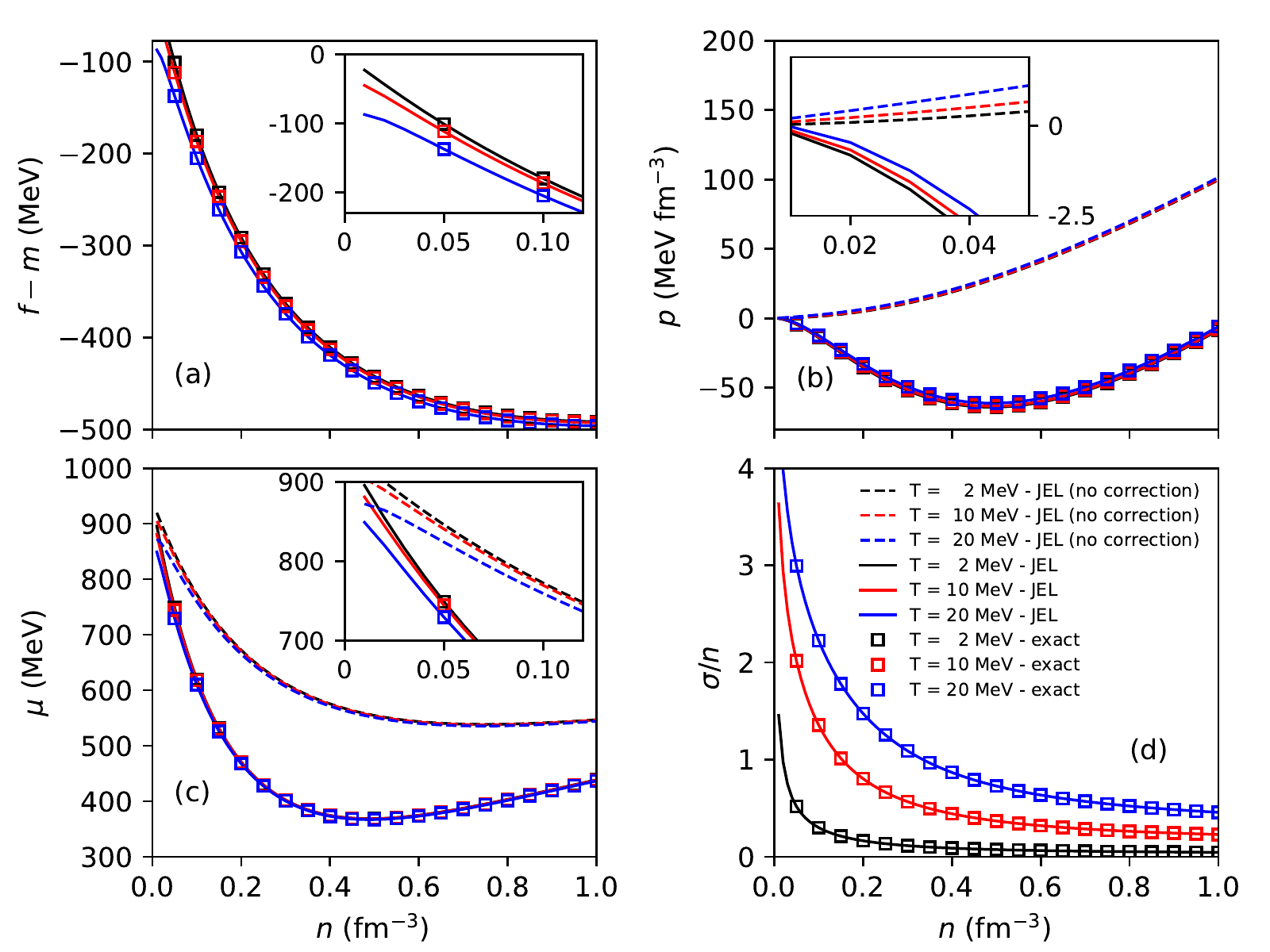}
\caption{Thermodynamic properties of the relativistic FG with density dependent effective mass in SM, as a function of the density and for the following set of temperature: $T=2$ (black), $10$ (red) and $20$~MeV (blue). The exact calculations are shown in squares while our formalism based on the JEL approximation is shown in solid lines. The different panels show the following quantities: (a) the Helmholtz free energy per particle $f$, (b) the pressure $p$, (c) a chemical potential $\mu$, and (d) the entropy per particle $\sigma/n$. Dashed lines represent the JEL approximation alone, i.e., disregarding the correction terms due to the density dependence of the effective mass.} 
\label{fig:rlsym}
\end{figure*}

We display in Fig.~\ref{fig:rlsym} the thermodynamical quantities described above, with the in-medium effective mass $m^*$ from the SLy5 parametrization~\citep{sly5}, by using both the JEL approximation (order 3) and the exact calculation, see Sec.~\ref{sec:exact}. 
For the former, besides $p^*$, $\epsilon^*$, and $\mu^*$ already mentioned, we also consider $\sigma^*$ and $n^*$, given in Eqs.~\eqref{eq:nodim} with constant $m^*$. We obtain a very good agreement between the JEL approximation and the exact calculation. This agreement is evaluated in a systematical way and given in Table~\ref{tab:chi-sym} where the residual difference $\xi_X$ is obtained from
\begin{equation}
\xi_X = \sqrt{\frac{1}{N}\sum_{i=1}^{N}\left(\frac{X_{\jel, i} - X_{\mbox{\tiny{exact}},i}}{X_{\mbox{\tiny{exact}},i}}\right)^2}, 
\label{eq:chi}
\end{equation}

with $N=20$ as in Fig.~\ref{fig:rlsym}. Table~\ref{tab:chi-sym} shows that the JEL approximation gives highly accurate results for all the densities and temperatures which are plotted in Fig.~\ref{fig:rlsym}. The deviation is almost independent of the temperature and is similar to the JEL approximation itself, see Ref.~\cite{JEL-1996-1020}.

\begin{table}[tb]
\centering
\tabcolsep=0.35cm
\caption{Comparison between the exact calculations and the JEL approximation, defined as the residual difference~$\xi_X$~\eqref{eq:chi} obtained for the thermodynamical quantities shown in Fig.~\ref{fig:rlsym} as a function of the temperature and using 20 densities points.}
\begin{tabular}{ccccc}
\hline\hline
$T$   & $\xi_{f - m}$ & $\xi_p$  & $\xi_{\mu}$ & $\xi_{\sigma/n}$ \\ 
(MeV) & ($10^{-6}$)  & ($10^{-4}$) & ($10^{-5}$)  &  ($10^{-3}$)\\ 
\hline    
$~~2$ & $4.76$ & $2.53$ & $1.80$ & $1.19$  \\ 
$10$  & $5.37$ & $3.07$ & $1.93$ & $0.26$ \\ 
$20$  & $5.78$ & $5.38$ & $2.58$ & $0.13$\\ 
\hline
\end{tabular}
\label{tab:chi-sym}
\end{table}

In Fig.~\ref{fig:rlsym} we also show in dashed lines the pressure $p^*$ and the chemical potential $\mu^*$ obtained from the JEL approximation, disregarding $p_{\corr}$ and $\mu_{\corr}$. It is shown that $\mu$ and $p$ are very different from $\mu^*$ and $p^*$.
This illustrates the importance to consider the correction due to the in-medium effective mass.

\subsection{Asymmetric nuclear matter}
\label{sec:ANM}

\begin{figure*}[tb]
\centering
\includegraphics[scale=1.0]{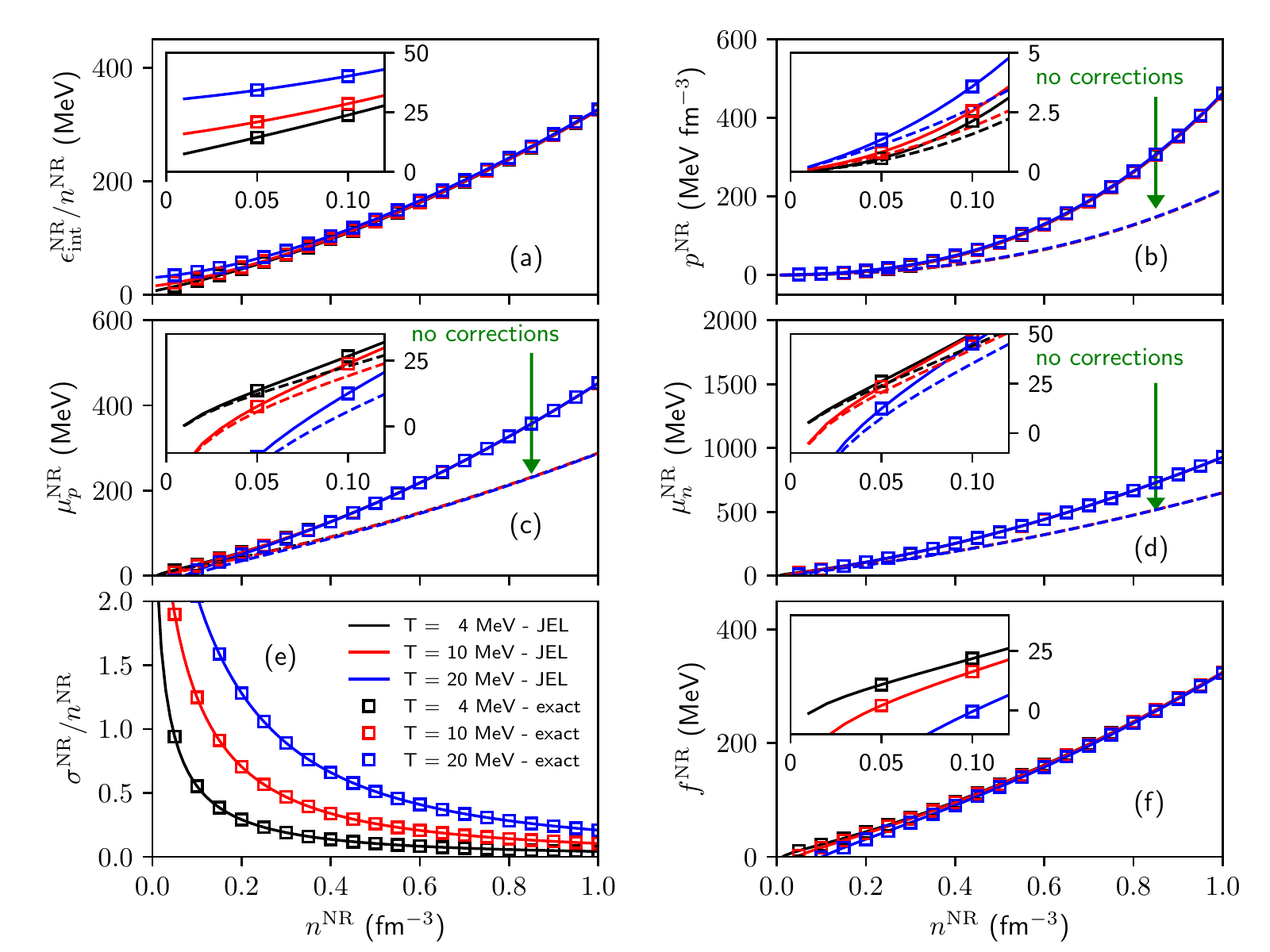}
\caption{Thermodynamic quantities of the non-relativistic FG, with density dependent effective mass, exact calculation (squares) and JEL approximation (full lines), for different temperatures and $\delta=0.4$: (a) energy per particle $\epsilon_{\rm int}^{\nrel}/n^{\nrel}$, (b) pressure $p^{\nrel}$, (c) proton chemical potential $\mu_p$, (d) neutron chemical potential $\mu_n$, (e) entropy per particle $\sigma^{\nrel}/n^{\nrel}$, and (f) Helmholtz free energy per particle $f^{\nrel}$. Dashed lines represent the JEL approximation without the correction terms due to the density dependence of the effective mass.} 
\label{fig:nrasym}
\end{figure*}

Now we generalize the expressions given in Sec.~\ref{sec:SNM} to isospin asymmetric matter (AM), i.e., to the case where the number of protons is different from the number of neutrons ($\delta\ne 0$). One, therefore, attributes a different in-medium effective mass to the neutrons and to the protons according to their respective density and to the global isospin asymmetry, as given from Eq.~\eqref{eq:effmass:den}.

The Helmholtz free energy density in AM reads
\begin{equation}
\phi = \phi(m^*_n,m^*_p,T) = \phi_n(m^*_n,T) + \phi_p(m^*_p,T) \, . 
\end{equation}

Following the steps given in Sec.~\ref{sec:SNM}, the pressure of the asymmetric system is obtained as
\begin{align}
p &\equiv n^2 \frac{\partial f}{\partial n}\Bigg|_{T,\delta} = \sum_{q=n,p} p^*_q+ p_{\corr,q}\, ,
\label{eq:press-asym}
\end{align}
with 
\begin{equation}
p^*_q = n_q\frac{\partial \phi_q}{\partial n_q}\Bigg|_{T,m^*_q,\delta} - \phi_q
\end{equation}
and 
\begin{equation}
p_{{\corr}, q} = n\frac{\partial \phi}{\partial m^*_q}\Bigg|_{T,n,\delta} \frac{\partial m^*_q}{\partial n}\Bigg|_{T,\delta} \, .
\label{eq:pcorr-asym}
\end{equation}
Note that in the expressions in AM, a quantity at constant $m^*$ ($\vert_{m^*}$) means constant $m_n^*$ and $m_p^*$, and constant $n$ ($\vert_{n}$) means constant $n_n$ and $n_p$.

The neutrons and protons chemical potentials are defined as,
\begin{equation}
\mu_q \equiv \frac{\partial \phi}{\partial n_q}\Bigg|_{T,n_{\bar{q}}} = \mu^*_q + \mu_{{\corr},q}\,,
\label{eq:muq}
\end{equation}
where $n_{\bar{q}}$ describes the other particles: neutrons if $q=p$ and protons otherwise. The correction terms to the chemical potential for the neutrons and the protons are given, respectively by
\begin{equation}
\mu_{{\corr},n} = \sum_{q=n,p}\frac{\partial \phi}{\partial m^*_q}\Bigg|_{T,n}\frac{\partial m^*_q}{\partial n_n}\Bigg|_{T,n_p}\, ,
\label{eq:muncorr}
\end{equation}
and
\begin{equation}
\mu_{{\corr},p} = \sum_{q=n,p}\frac{\partial \phi}{\partial m^*_q}\Bigg|_{T,n}\frac{\partial m^*_q}{\partial n_p}\Bigg|_{T,n_n} \, .
\label{eq:mupcorr}
\end{equation}

The nucleon chemical potential is determined from $\mu_p$ and $\mu_n$ as
\begin{align}
\mu &= \left(\frac{1-\delta}{2}\right)\mu_p + \left(\frac{1+\delta}{2}\right)\mu_n
\nonumber\\
&= \mu^* + \left(\frac{1-\delta}{2}\right)\mu_{{\corr},p} + \left(\frac{1+\delta}{2}\right)\mu_{{\corr},n},
\end{align}
with 
\begin{align}
\mu^*  &= \left(\frac{1-\delta}{2}\right)\mu_p^* + \left(\frac{1+\delta}{2}\right)\mu_n^*.
\end{align}

The entropy density $\sigma$ is not impacted by the density dependence of $m^*_q$ since it is defined as a derivative with respect to the temperature. In this case, Eq.~\eqref{eq:ssjel} still applies. 

We can determine the energy density in AM from the Euler relation, 
\begin{align}
\epsilon &= - p + \mu_p n_p + \mu_n n_n + T\sigma 
\nonumber\\
&= - p^* +\mu^* \, n + T\sigma^* 
= \epsilon^*,
\label{eq:eden-asym}
\end{align}
i.e., the energy density is not modified by the density dependence of the effective mass even in AM. 
In Eq.~\eqref{eq:eden-asym}, $p^*=p^*_n+p^*_p$.

For the Helmholtz free energy density, it is possible to perform the same analysis as the one done in Sec.~\ref{sec:SNM} to obtain Eq.~\eqref{eq:freeenergy}, and for the density itself ($n=n^*$). More details are given in Appendix~\ref{app:asymexp}.

We test in AM the goodness of our formalism based on the JEL approximation and compare our results to the exact calculations in Fig.~\ref{fig:nrasym}, considering $\delta=0.4$ and $\gamma=2$. We focus on the non-relativistic (NR) FG and we consider the effective mass from the SLy5 parametrization. The NR pressure and energy density are defined as,
\begin{align}
p_q^{\nrel} &= \frac{\gamma}{6\pi^2}\int_0^{\infty}\hspace{-0.2cm}dk\,k^2\frac{k^2}{m_q}F_D(k,\mu,T,m_q)\, ,
\label{pressfg-nr}
\end{align}
and
\begin{align}
\epsilon_q^{\nrel} &= \frac{\gamma}{2\pi^2}\int_0^{\infty}\hspace{-0.2cm}dk\,k^2\frac{k^2}{2m_q}F_D(k,\mu,T,m_q)\, ,
\label{eq:edenfg-nr}
\end{align}
satisfying the expected relation $\epsilon^{\nrel}=\frac 3 2 p^{\nrel}$ for FFG and FG systems.

\begin{table*}[tb]
\centering
\tabcolsep=0.68cm
\caption{Comparison between the exact calculations and the JEL approximation based on the $\xi_X$ analysis for the non-relativistic thermodynamical quantities shown in Fig.~\ref{fig:nrasym}.}
\begin{tabular}{ccccccc}
\hline\hline
$T$   & $\xi_{\epsilon^{\nrel}/n}$ & $\xi_{p^{\nrel}}$  & $\xi_{\mu^{\nrel}_p}$ & $\xi_{\mu^{\nrel}_n}$ & $\xi_{\sigma^{\nrel}/n}$ & $\xi_{f^{\nrel}}$\\ 
(MeV) & ($10^{-4}$)  & ($10^{-4}$) & ($10^{-4}$) &  ($10^{-4}$) & ($10^{-3}$) &($10^{-4}$)\\ 
\hline    
$~~4$ & $1.21$  & $2.90$ & $1.56$ & $2.30$ & $3.61$ & $1.23$\\
$10$  & $1.23$  & $3.10$ & $1.73$ & $2.45$ & $0.45$ & $1.67$\\
$20$  & $1.21$  & $3.09$ & $2.49$ & $2.49$ & $0.39$ & $4.53$\\
\hline
\end{tabular}
\label{tab:chi-asym}
\end{table*}

In order to compute the non-relativistic JEL expressions from the relativistic ones given in Ref.~\cite{JEL-1996-1020}, we use the procedure described in Appendix~\ref{app:jelnr}, by taking Eqs.~\eqref{eq:pjelnr}, \eqref{eq:edjelnr}-\eqref{eq:mujelnr} for each component of the system (protons and neutrons). Similarly to SM, we verify the excellent agreement between exact calculations and JEL approximation complemented by the in-medium effective mass. The $\xi_X$ reported in Table~\ref{tab:chi-asym} are compatible with the ones given in Ref.~\cite{JEL-1996-1020}, showing that the main origin of the small difference is mostly due to the JEL approximation itself. Moreover, notice that, as in the previous case, the computation of pressure and chemical potentials without the correction terms, in this case, given by Eqs.~\eqref{eq:pcorr-asym}, \eqref{eq:mupcorr}, and \eqref{eq:muncorr}, leads to a mismatch of the curves displayed in panels (b), (c), (d).

\section{In-medium mass depending on density and temperature}
\label{sec:effmass-nt}

We now consider a more general case where the in-medium effective mass depends on the density and on the temperature, see Sec.~\ref{sec:examples}. We have $m_q^*=m_q^*(n,\delta,T)$, or equivalently, $m_q^*=m_q^*(n_p,n_n,T)$. For this case, pressure and chemical potentials remain unchanged because they are obtained from derivatives taken at a constant temperature. Therefore, they are still described by Eqs.~\eqref{eq:press-asym}, and~\eqref{eq:muq}. The same is not true for quantities defined as derivatives with respect to the temperature, since the following generic equation holds
\begin{align}
\frac{\partial X(n,T,m^*)}{\partial T}\Bigg|_{n} &= \frac{\partial X(n,T,m^*)}{\partial T}\Bigg|_{n,m^*} 
\nonumber\\
&+ \frac{\partial X(n,T,m^*)}{\partial m^*}\Bigg|_{n,T} \frac{\partial m^*}{\partial T}\Bigg|_{n},
\label{eq:general-t}
\end{align}
i.e, there is now an additional correction, $\partial m^*/\partial T$, originating from the temperature dependence of $m^*$. The entropy is impacted by this new term as,
\begin{align}
\sigma &= -\frac{\partial\phi}{\partial T}\Bigg|_{n} \nonumber\\
&= - \sum_{q=p,n}\left( \frac{\partial \phi^*_q}{\partial T}\Bigg|_{n,m^*} 
+ \frac{\partial\phi^*_q}{\partial m^*_q}\Bigg|_{n,T} \frac{\partial m^*_q}{\partial T}\Bigg|_{n}  \right)
\nonumber\\
&= \sum_{q=p,n} \left(\sigma^*_q + \sigma_{\corr,q} \right) \, ,
\end{align}
with
\begin{align}
\sigma^*_{q} = -\frac{\partial \phi^*_q}{\partial T}\Bigg|_{n,m^*} \,, \hbox{ and }
\sigma_{\corr,q} = -\frac{\partial\phi^*_q}{\partial m^*_q}\Bigg|_{n,T} \frac{\partial m^*_q}{\partial T}\Bigg|_{n} ,
\end{align}
i.e., an explicit correction has to be taken into account for the calculation of the entropy. We define $\sigma^*=\sigma^*_n+\sigma^*_p$.

The energy density of the system is also affected by this correction, as one can verify:
\begin{align}
\epsilon &= \phi + T\sigma = \phi^* + T\sigma^* + T\sum_{q=p,n}\sigma_{\corr,q}
\nonumber\\
&=\epsilon^* + \epsilon_{\corr},
\end{align}
with 
\begin{align}
\epsilon_{\corr} = T\sum_{q=p,n}\sigma_{\corr,q} \,.
\end{align}

In summary, the temperature dependence of the effective mass impacts the calculation of the entropy and of the energy density.

\section{In-medium mass depending on the chemical potential}
\label{sec:effmass-cp}

We now consider the case where the in-medium mass depends on the chemical potential, and for simplicity, we consider SM. The natural ensemble is the GCE with the potential $\Omega=\Omega(\mu,T,m^*(\mu))=-p(\mu,T,m^*(\mu))$. 

It is worth noting that the functional form of $\Omega$ is not modified by the chemical dependency of the effective mass. The only change is the replacement of $m$ by $m^*$.
As a consequence, the pressure satisfies $p=p^*$ without correction induced by the in-medium effective mass.

In general, one has 
\begin{align}
\frac{\partial X(\mu,T,m^*)}{\partial \mu}\Bigg|_{T}  &= \frac{\partial X(\mu,T,m^*)}{\partial \mu}\Bigg|_{T,m^*} 
\nonumber\\
&+ \frac{\partial X(\mu,T,m^*)}{\partial m^*}\Bigg|_{T,\mu} \frac{\partial m^*}{\partial \mu}\Bigg|_{T},
\label{general-gce}
\end{align}
for a thermodynamical quantity $X$.
The density of the system is obtained through
\begin{eqnarray}
n &\equiv& -\frac{\partial \Omega(\mu,T,m^*)}{\partial \mu}\Bigg|_{T} = \frac{\partial p}{\partial \mu}\Bigg|_{T,m^*} + \frac{\partial p}{\partial m^*}\Bigg|_{T,\mu} \frac{\partial m^*}{\partial \mu}\Bigg|_{T} \nonumber \\
&=& n^* + n_{\corr},
\end{eqnarray}
with
\begin{equation}
n_{\corr} = \frac{\partial p}{\partial m^*}\Bigg|_{T,\mu} \frac{\partial m^*}{\partial \mu}\Bigg|_{T},
\end{equation}
and $n^*$ being obtained from Eq.~\eqref{eq:nodim} with $m$ replaced by the constant $m^*(\mu)$. Since the entropy density is defined from a derivative of the temperature, it is not impacted by the in-medium effective mass and we have,
\begin{equation}
\sigma=\sigma^*.
\end{equation}
For the energy density $\epsilon$, we have
\begin{align}
\epsilon &= -p + \mu\, n + T\sigma = -p^*  + \mu (n^* + n_\corr) + T\sigma^*
\nonumber\\
&= \epsilon^* + \epsilon_{\corr}
\end{align}
with
\begin{equation}
\epsilon_{\corr} = \mu \, n_{\corr},
\end{equation}
and for the Helmholtz free energy density, one has
\begin{align}
\phi &= \epsilon - T\sigma = \epsilon^* - T\sigma^* + \epsilon_{\corr} \nonumber \\
&= \phi^* + \epsilon_{\corr},
\end{align}
i.e., there are corrections in $\epsilon$ and $\phi$ given explicitly in term of the derivative of $m^*$ with respect to $\mu$.

\section{Conclusions}
\label{sec:conclusions}

In this paper, we have shown how approximations for the FFG at finite-$T$ and with arbitrary masses can be employed in the more general case of the FG with in-medium effective masses. For density dependent effective masses, we have performed a detailed comparison of the exact calculation of the thermodynamical quantities by calculating the Fermi integral with a very efficient approximation for the FFG, here we considered the JEL approximation at order 3 from Ref.~\cite{JEL-1996-1020}. The comparison has shown an excellent agreement between the exact calculations and our formalism based on the JEL approximation. The residual difference being mostly due to the JEL approximation itself, and not to our formalism.

We have also considered other dependencies of the in-medium effective mass, such as a dependency on the density and temperature, or a dependency on the chemical potential. In all cases, simple relations can be derived and generalized as the approximations for the FG.

Our expression can be safely employed in astrophysics where numerical approximations of the FG are useful to calculate the kinetic energy contribution with in-medium effective masses. It keeps the rapidity of analytical expressions for the calculation of finite-temperature thermodynamical quantities and can be employed to calculation of finite temperature equation of states necessary for instance, for neutron star mergers or core-collapse supernovae.

\begin{acknowledgments}
This work is a part of the project INCT-FNA proc. No. 464898/2014-5. It is also supported by Conselho Nacional de Desenvolvimento Cient\'ifico e Tecnol\'ogico (CNPq) under Grants No. 312410/2020-4 (O.~L.), and No. 308528/2021-2 (M.~D.). O.~L., and M.~D. also acknowledge Funda\c{c}\~ao de Amparo \`a Pesquisa do Estado de S\~ao Paulo (FAPESP) under Thematic Project 2017/05660-0. O.~L. is also supported by FAPESP under Grant No. 2022/03575-3 (BPE). This study was financed in part by the Coordena\c{c}\~ao de Aperfei\c{c}oamento de Pessoal de N\'ivel Superior - Brazil (CAPES) - Finance Code 001 - Project number 88887.687718/2022-00 (M.~D.).
J.M. is supported by the CNRS-IN2P3 MAC masterproject, and benefit from PHAROS COST Action MP16214 and the LABEX Lyon Institute of Origins (ANR-10-LABX-0066) of the \textsl{Universit\'e de Lyon} for its financial support within the program \textsl{Investissements d'Avenir} (ANR-11-IDEX-0007) of the French government operated by the National Research Agency (ANR).
\end{acknowledgments}

\appendix
\section{Numerical approximation at finite temperature with constant mass particles.}
\label{app:jel}

Here we present the JEL expressions given in Ref.~\cite{JEL-1996-1020}. These expressions are used for the numerical calculations presented in this paper.

\subsection{General relativistic framework}
\label{app:rljel}

In the following, the following dimensionless thermodynamical quantities are introduced
\begin{align}
p^* &= d \bar{p} \, ,
\epsilon^* = d \bar{\epsilon} \, ,
\phi^* = d \bar{\phi}  \, , \nonumber \\
\sigma^* &= \frac{d}{m^*} \bar{\sigma} \, ,
n^* = \frac{d}{m^*} \bar{n}\, ,
\label{eq:nodim}
\end{align}
where $d=\gamma m^{*4}/(2\pi^2)$. Along the entire appendix, $m^*$ is the constant Fermion mass, and therefore, does not depend on density, temperature, or chemical potential.

The dimensionless pressure $\bar{p}$ is given by the following approximation:
\begin{equation}
\bar{p} = \frac{fg^{5/2}(1+g)^{3/2}}{(1+f)^{J+1}(1+g)^L} \sum_{j=0}^J \sum_{l=0}^L p_{jl}f^jg^l,
\label{eq:dpjel}
\end{equation}
where $g = t(1 + f)^{1/2}$ and $t = T/m^*$, with $T$ being the temperature. The chemical potential reads
\begin{equation}
\mu^* = \psi T + m^*,
\label{eq:mujel}
\end{equation}
with the function $\psi$ given by
\begin{equation}
\psi = 2\sqrt{1 + \frac{f}{a}} + \ln{\frac{\sqrt{1 + f/a}-1}{\sqrt{1 + f/a}+1}}. 
\label{eq:psijel}
\end{equation}
The constants $p_{jl}$ and $a$ are given in Table~\ref{tab:const} for the approximation of order 3. 

\begin{table}[tb]
\centering
\tabcolsep=0.2cm
\caption{Coefficients for the pressure of Fermions considering $J = L = 3$ and $a = 0.433$ extracted from Ref.~\cite{JEL-1996-1020}.}
%\begin{ruledtabular}
\begin{tabular}{cDDDD}
\hline\hline
\decimals  
$p_{jl}$  & l = 0  & l = 1 & l = 2 & l = 3 \\ 
\hline    
  $j = 0$   &  5.34689 & 18.0517 & 21.3422 &  8.53240 \\ 
  $j = 1$   & 16.8441  & 55.7051 & 63.6901 & 24.6213  \\ 
  $j = 2$   & 17.4708  & 56.3902 & 62.1319 & 23.2602  \\ 
  $j = 3$   &  6.07364 & 18.9992 & 20.0285 &  7.11153 \\
\hline
\end{tabular}
%\end{ruledtabular}
\label{tab:const}
\end{table}

The dimensionless density $\bar{n}$ and the internal energy density, $\bar{\epsilon}_\intern$, are expressed, respectively, as
\begin{align}
\bar{n} &= \frac{f\left[g(1+g)\right]^{3/2}}{(1+f)^{J+1/2}(1+g)^L\sqrt{1+f/a}} \times \nonumber \\ 
&\times \sum_{j=0}^J \sum_{l=0}^L p_{jl}f^jg^l\left[1 + l + \left(\frac{1}{4} + \frac{l}{2}-J\right)\frac{f}{1+f} + \right. \nonumber \\ 
&+ \left.\left(\frac{3}{4} - \frac{L}{2}\right)\frac{fg}{(1+f)(1+g)} \right],
\label{eq:dnjel}
\end{align}
and
\begin{align}
\bar{\epsilon}_\intern &= \frac{fg^{5/2}(1+g)^{3/2}}{(1+f)^{J+1}(1+g)^L} \times \nonumber \\
&\times \sum_{j=0}^J \sum_{l=0}^L p_{jl}f^jg^l\left[\frac{3}{2} + l + \left(\frac{3}{2} - L\right)\frac{g}{1+g}\right] .
\label{eq:diejel}
\end{align} 
From Eqs.~\eqref{eq:dnjel} and~\eqref{eq:diejel} one can define dimensionless energy density as
\begin{equation}
\bar{\epsilon} = \bar{\epsilon}_\intern + \bar{n} \,.
\end{equation}
The dimensionless entropy and Helmholtz free energy density can be found, respectively, as
\begin{equation}
t\bar{\sigma} = \bar{\epsilon} + \bar{p} - (\psi t + 1) \bar{n} = \bar{\epsilon}_\intern + \bar{p} - \bar{n} \psi t
\end{equation}
and
\begin{equation}
\bar{\phi} = \bar{\epsilon} - t\bar{\sigma} = -\bar{p} + (\psi t + 1) \bar{n} \,.  
\end{equation}

From Eqs.~\eqref{eq:dpjel} and~\eqref{eq:diejel}, it is possible to obtain an analytical expression for the derivative of the pressure:
\begin{align}
\frac{\partial p^*}{\partial m^*}\Bigg|_{n,T} &= \frac{3}{2}\frac{d}{m^*}\bar{p} 
\nonumber \\
&- \frac{d}{m^*}\frac{fg^{5/3}(1+g)^{3/2}}{(1+f)^{J+1}(1+g)^L} \sum_{j=0}^J \sum_{l=0}^L lp_{jl}f^jg^l \times
\nonumber\\
&\times \left[l + \left(\frac{3}{2} - L\right)\frac{g}{1+g}\right] 
\nonumber \\
&= \frac{3}{2}\frac{d\, \bar{p}}{m^*} - \frac{d}{m^*}\left(\bar{\epsilon}_\intern - \frac{3}{2}\bar{p}\right)  
\nonumber \\
&= \frac{1}{m^*}(3p^*-\epsilon^*) + n^*\, ,
\label{eq:dpjeldm}
\end{align}
which is useful to calculate $\partial\phi^*/\partial m^*$ in Sec.~\ref{sec:JEL}.

\subsection{Non-relativistic reduction of JEL expressions}
\label{app:jelnr}

It is possible to deduce the non-relativistic expressions of the thermodynamical quantities from their relativistic JEL approximation. To do so, we inject in Eqs.~\eqref{eq:dpjel}, \eqref{eq:dnjel}, and \eqref{eq:diejel} an artificial large mass $m_{\lar}c^2\gg k_F c$ instead of $m^*$. In this way, the thermodynamical quantities reach the non-relativistic limit, but for a mass that is not one of the particles. In order to recover the proper mass, we perform a leading order analysis of the thermodynamical quantities, with $t,g\rightarrow 0$, from which we deduce the correct mass ratio $m_{\lar}/m^*$. This ratio is defined as the one which allows us to recover the non-relativistic expression. Note that the non-relativistic expression is independent of the value taken for $m_{\lar}$, provided it is large enough.

For instance, considering a large mass $m_\lar$ we obtain for the pressure
\begin{align}
\bar{p}_{\lar}\sim g^{5/2}\sim m_{\lar}^{-5/2}\, ,
\end{align}
from which we deduce the non-relativistic pressure as
\begin{align}
p^{\nrel*} = d\bar{p}^{\nrel}
\label{eq:pjelnr}
\end{align}
with
\begin{align}
\bar{p}^{\nrel}=\left(\frac{m_{\lar}}{m^*}\right)^{5/2}\bar{p}_{\lar}.
\end{align}
Similarly, for the density and the internal energy, we obtain
\begin{align}
\bar{n}_{\lar}&\sim g^{3/2}\sim m_{\lar}^{-3/2},
\\
\bar{n}^{\nrel}&=\left(\frac{m_{\lar}}{m^*}\right)^{3/2}\bar{n}_{\lar}, 
\end{align}
and 
\begin{align}
\bar{\epsilon}_{\intern,\lar}&\sim g^{5/2}\sim m_{\lar}^{-5/2},
\\
\bar{\epsilon}_\intern^{\nrel}&=\left(\frac{m_{\lar}}{m^*}\right)^{5/2}\bar{\epsilon}_{\intern,\lar}\, .
\end{align}
Concerning energy density, its non-relativistic expression is obtained as 
\begin{align}
\epsilon_{\rm int}^{\nrel*}=d(\bar{\epsilon}_\intern^{\nrel} + \bar{n}^{\nrel}) - m^* n^{\nrel*},
\label{eq:edjelnr}
\end{align}

and the entropy density is given by
\begin{align}
\sigma^{\nrel *} = \frac{d}{m^* t}(\bar{\epsilon}_\intern^{\nrel} + \bar{p}^{\nrel} - \bar{n}\psi t) \, .
\label{eq:sjelnr}
\end{align}
The Helmholtz free energy density is expressed as
\begin{align}
\phi^{\nrel *}=d[-\bar{p}^{\nrel} + (\psi t+1)\bar{n}^{\nrel}] - m^* n^{\nrel *},
\label{eq:fjelnr}
\end{align}
and the chemical potential is written as
\begin{equation}
\mu^{\nrel *} = \psi T \, .
\label{eq:mujelnr}
\end{equation}

\section{Isospin asymmetric matter}
\label{app:asymexp}

We consider density dependent in-medium effective mass as in Sec.~\ref{sec:ANM}. The total pressure is defined as
\begin{align}
p &= n\frac{\partial \phi}{\partial n}\Bigg|_{T} - \phi 
\nonumber\\
&= n\left(\frac{\partial \phi}{\partial n}\Bigg|_{T,m^*} + \sum_{q=n,p}\frac{\partial \phi}{\partial m^*_q}\Bigg|_{T,n}\frac{\partial m^*_q}{\partial n}\Bigg|_{T} \right)- \phi
\nonumber \\
&= \sum_{q=n,p}\left(n_q\frac{\partial \phi_q}{\partial n_q}\Bigg|_{T,m^*_q} - \phi_q\right) 
\nonumber\\
&+ n\sum_{q=n,p}{\frac{\partial \phi}{\partial m^*_q}\Bigg|_{T,n}\frac{\partial m^*_q}{\partial n}}\Bigg|_{T} \nonumber \\
&= p^* + p_{\corr}\, .
\end{align}
For the protons' chemical potential, we have
\begin{align}
\mu_p &= \frac{\partial \phi}{\partial n_p}\Bigg|_{T,m^*,n_n} +
\frac{\partial \phi}{\partial m^*_p}\Bigg|_{T,n}\frac{\partial m^*_p}{\partial n_p}\Bigg|_{T,n_n} 
\nonumber\\
&+ \frac{\partial \phi}{\partial m^*_n}\Bigg|_{T,n}\frac{\partial m^*_n}{\partial n_p}\Bigg|_{T,n_n} 
\nonumber \\
&= \mu^*_p + \mu_{{\corr},p},
\end{align}
and similarly for the neutrons.

The energy density of the asymmetric system can be determined from the Euler relation, namely,
\begin{align}
\epsilon &=  - p + \mu_p n_p + \mu_n n_n + T\sigma
\nonumber\\
&= - p^* +\mu^*n + T\sigma^* - \sum_{q=n,p}(p_{{\corr},q}-\mu_{{\corr},q}\, n_q) \, .
\label{eq:eden}
\end{align}
The use of Eqs.~\eqref{eq:pcorr-asym}, \eqref{eq:mupcorr}, and \eqref{eq:muncorr} into Eq.~\eqref{eq:eden}, along with 
\begin{align}
\frac{\partial m^*_p}{\partial n_p} &= \frac{\partial m^*_p}{\partial n} + \frac{\partial m^*_p}{\partial \delta}\frac{\partial \delta}{\partial n_p}, 
\\
\frac{\partial m^*_p}{\partial n_n} &= \frac{\partial m^*_p}{\partial n} + \frac{\partial m^*_p}{\partial \delta}\frac{\partial \delta}{\partial n_n}, 
\\
\frac{\partial m^*_n}{\partial n_p} &= \frac{\partial m^*_n}{\partial n} + \frac{\partial m^*_n}{\partial \delta}\frac{\partial \delta}{\partial n_p}, 
\\
\frac{\partial m^*_n}{\partial n_n} &= \frac{\partial m^*_n}{\partial n} + \frac{\partial m^*_n}{\partial \delta}\frac{\partial \delta}{\partial n_n}, 
\end{align}
and
\begin{align}
\frac{\partial \delta}{\partial n_p} = -\frac{2n_n}{n^2},\qquad 
\frac{\partial \delta}{\partial n_n} = \frac{2n_p}{n^2},
\end{align}
leads to
\begin{align}
\epsilon =  - p^* +\mu^* \, n + T\sigma^* = \epsilon^*,
\end{align}
showing that no corrections for the energy density due to the density dependence of the effective mass are needed in an asymmetric system.

\bibliography{references}{}
\bibliographystyle{aasjournal}

\end{document}